\documentclass[final,5p,times,twocolumn]{elsarticle}
\biboptions{sort&compress}
\usepackage{amsmath,amssymb}

\journal{Physics Letters B}

\usepackage{graphicx}
\usepackage{color}
\usepackage[textsize=tiny]{todonotes}

\usepackage{picture}
\usepackage{calc}
\usepackage{subfigure}

\usepackage{color}

\usepackage{tikz}
\usetikzlibrary{shapes,arrows,automata,positioning}
\usepackage{tikz-3dplot}

\usepackage{comment}

\usepackage[bookmarks=true,hyperfigures=true]{hyperref}
\usepackage{fixmath}
\usepackage{mathtools}

\providecommand{\href}[2]{#2}

\let\oldbfseries=\bfseries
\let\oldmdseries=\mdseries
\let\oldnormalfont=\normalfont
\renewcommand{\bfseries}{\oldbfseries\boldmath}
\renewcommand{\mdseries}{\oldmdseries\unboldmath}
\renewcommand{\normalfont}{\oldnormalfont\unboldmath}

\makeatletter
\newlength{\apb@width}
\newcommand{\autoparbox}[2][c]{\settowidth{\apb@width}{#2}\parbox[#1]{\apb@width}{#2}}

\makeatother

\RequirePackage{verbatim}

\makeatletter
\newwrite\bibinl@out
{\immediate\closeout\bibinl@out\@esphack}
\makeatother

\newcommand{\e}{\operatorname{e}}
\newcommand{\de}{\operatorname{d}\!}
\DeclareMathOperator{\tr}{tr}

\newcommand{\eqndot}{\, . }
\newcommand{\eqncom}{\, , }
\newcommand{\eqnsem}{\, ; }
\DeclareMathOperator{\cder}{D}
\newcommand{\scal}{\phi}
\newcommand{\scalc}{\scal^{\text{cl}}}
\newcommand{\scalq}{\tilde{\scal}}
\newcommand{\ferm}{\psi}
\newcommand{\aferm}{\bar{\ferm}}
\newcommand{\comm}[2]{[#1,#2]}

\DeclareMathOperator{\phaneq}{\phantom{{}=}}

\newcommand{\SU}[1]{\operatorname{SU}(#1)}

\newcommand{\YM}{{\mathrm{\scriptscriptstyle YM}}}

\newcommand{\combi}{\eta}

\begin{document}

\begin{frontmatter}

\title{One-loop Wilson loops and the particle-interface potential in AdS/dCFT}

\author{Marius de Leeuw, Asger C. Ipsen, Charlotte Kristjansen and Matthias Wilhelm}
\address{Niels Bohr Institute, Copenhagen University,\\ Blegdamsvej 17, 2100 Copenhagen \O{}, Denmark}
\ead{deleeuwm@nbi.ku.dk, asgercro@nbi.ku.dk, kristjan@nbi.ku.dk, matthias.wilhelm@nbi.ku.dk}

\begin{abstract}
We initiate the calculation of quantum corrections to Wilson loops in a class of four-dimensional defect conformal field theories with vacuum expectation values based on $\mathcal{N}=4$ super Yang-Mills theory.
Concretely, we consider an infinite straight Wilson line, obtaining explicit results for the one-loop correction to its expectation value in the large-$N$ limit.
This allows us to extract the particle-interface potential of the theory.
In a further double-scaling limit, we compare our results to those of a previous calculation in the dual string-theory set-up consisting of a D5-D3 probe-brane
system with flux, and we find perfect agreement.
\end{abstract}


\end{frontmatter}

\section{Introduction}

Wilson loops form an important class of observables in any gauge theory.
Among others, they are related to scattering amplitudes \cite{Alday:2007hr} and can be used to determine the quark-antiquark potential \cite{Wilson:1974sk}.
In gauge theories that interact with a boundary or interface, a further important potential is the particle-interface potential, which can likewise be  obtained from a Wilson loop.
A simple set-up to study this potential is given by defect conformal field theories (dCFTs).

We will be interested in dCFTs with holographic duals, of which a number of examples exist, building on the Karch-Randall
idea~\cite{Karch:2000gx}. More precisely, we will consider ${\cal N}=4$ supersymmetric Yang-Mills theory ($\mathcal{N}=4$ SYM theory) with a codimension-one defect at $x_3=0$ separating two regions of space-time where the gauge group is $\SU{N}$ and $\SU{N-k}$, respectively \cite{Nahm:1979yw,Diaconescu:1996rk,Giveon:1998sr,Constable:1999ac}. 
To achieve the difference in the rank of the gauge group, a non-vanishing vacuum expectation value (vev) proportional to $1/x_3$ is
assigned to three of the scalar fields in the region $x_3>0$.
The resulting gauge theory is dual to a  D5-D3 probe-brane system involving a single D5 brane of geometry $AdS_4\times S^2$ supporting a background gauge field flux of $k$ units through the $S^2$.

A special feature of dCFTs is that one-point functions of gauge-invariant local composite operators can be non-vanishing \cite{Cardy:1984bb}. One-point functions of the present dCFT were studied at tree level in 
\cite{Nagasaki:2012re,Kristjansen:2012tn,deLeeuw:2015hxa,Buhl-Mortensen:2015gfd,deLeeuw:2016umh}. Furthermore,
in \cite{Buhl-Mortensen:2016pxs} we set up the program for performing perturbative calculations in the dCFT and treated
one-point functions as the first simple application.
This opened the possibility of performing more elaborate comparisons between gauge- and string-theory
results in a certain double-scaling limit, which is imposed on top of the planar limit     by taking the 't Hooft coupling $\lambda\to\infty$ and $k\to\infty$ while keeping $\lambda/k^2$ finite, as proposed in~\cite{Nagasaki:2011ue,Nagasaki:2012re}.

In this letter, we initiate the calculation of quantum corrections to non-local observables in the dCFT under consideration.
Concretely, we calculate the one-loop correction of a particular Wilson loop in the large-$N$ limit
which allows us to extract the corresponding correction to the particle-interface potential.
This Wilson loop was already considered at tree level in \cite{Nagasaki:2011ue}, where also the corresponding string-theory calculation was performed.
After giving a brief introduction to the dCFT in section~\ref{sec: defect theory}, we describe our calculation of the Wilson loop 
in section \ref{sec: Wilson loop}. This calculation builds heavily on \cite{Buhl-Mortensen:2016pxs}, and we refer the reader to this
reference and to the forthcoming article~\cite{longpaper} for details.
In section \ref{sec: comparison}, we then compare our planar result to the string-theory result of \cite{Nagasaki:2011ue} in the aforementioned double-scaling limit, 
finding perfect agreement 
exactly as for one-point functions~\cite{Buhl-Mortensen:2016pxs}. Finally, section~\ref{sec: conclusion} contains our conclusion and outlook.

\section{The Defect Theory}
\label{sec: defect theory}
The action of the dCFT is given by the action of the four-dimensional $\mathcal{N}=4$ SYM theory plus an action describing a three-dimensional fundamental hypermultiplet on the defect as well as its coupling to the bulk fields \cite{Nahm:1979yw,Diaconescu:1996rk,Giveon:1998sr,Constable:1999ac}.
We use the following form of the action of $\mathcal{N}=4$ SYM theory 
\begin{equation}
 \label{eq: SYM-action}
 \begin{aligned}
 S_{{\cal N}=4}&=\frac{2}{g_\YM^2}\int \de^4x\tr\,\biggl( -\frac{1}{4}F_{\mu\nu}F^{\mu\nu}-\frac{1}{2}\cder_\mu\scal_i\cder^\mu\scal_i\\
 &\phaneq+\frac{i}{2}\aferm\Gamma^\mu\cder_\mu\ferm +\frac{1}{2}\aferm\Gamma^i\comm{\scal_i}{\ferm}+\frac{1}{4}\comm{\scal_i}{\scal_j}\comm{\scal_i}{\scal_j}\biggr)\eqnsem
\end{aligned}
\end{equation}
see \cite{Buhl-Mortensen:2016pxs,longpaper} for details on our conventions.
As in the case of one-point functions, the three-dimensional action will not play any role in the present one-loop calculation; the reason is that the only potentially contributing Feynman diagram would involve a loop consisting of a single propagator
of a defect field which vanishes due to conformal invariance.\footnote{Defect fields can come into play at higher loop orders.}

Of the different fields in $\mathcal{N}=4$ SYM theory, three scalars acquire a non-vanishing vev encoding the so-called
fuzzy-funnel solution \cite{Constable:1999ac}:
\begin{equation}
\label{eq: classical solution}
\langle\phi_i\rangle_{\text{tree}}= \scalc_i=-\frac{1}{x_3} t_i\oplus 0_{(N-k)\times(N-k)}\eqncom \hspace{0.5cm} x_3>0 \eqncom
\end{equation}
where $i=1,2,3$.
Here, $t_1,t_2$ and $t_3$ are the generators of the $k$-dimensional irreducible representation of the $\SU{2}$ Lie algebra. 
In particular, $t_3$ is a diagonal matrix with eigenvalues 
\begin{equation}
\label{eq: eigenvalues of t3}
 d_{k,i}=\frac{1}{2}(k-2i+1)\eqncom \qquad i=1,\dots,k\eqndot
\end{equation}

In order to calculate quantum corrections, we expand the action \eqref{eq: SYM-action} around the classical solution, writing
\begin{equation}
\scal_i=\scalc_i+\scalq_i\eqncom \hspace{0.5cm} i=1,2,3\eqndot
\end{equation}
The explicit form of the (gauge-fixed) action resulting from this is given in \cite{Buhl-Mortensen:2016pxs,longpaper}.
In contrast to the usual action \eqref{eq: SYM-action} of $\mathcal{N}=4$ SYM theory, this action contains a quadratic mass-like term, which moreover has two non-standard properties.

The first non-standard property of the mass-like term is that it is non-diagonal in the colour components of the different fields. This is caused by the classical solution \eqref{eq: classical solution} taking values in the $k$-dimensional representation of the Lie algebra $\SU{2}$ in colour space. Moreover, also some of the flavours mix, concretely the scalars $\scalq_1$, $\scalq_2$ and $\scalq_3$ with the component $A_3$ of the gauge field and the fermion flavours among each other.
This mixing problem was solved in \cite{Buhl-Mortensen:2016pxs,longpaper}. The eigenvalues are given in table~\ref{tab:spectrum}, partially in terms of  
\begin{equation}
\label{eq: def nu}
 \nu=\sqrt{m^2+\frac{1}{4}} \eqndot
\end{equation}

\setcounter{footnote}{0}

\begin{table}[t]
\begin{tabular}{c|c|c|c}
Multiplicity & \!$\nu(\scalq_{4,5,6},A_{0,1,2},c)$\! & \!$m(\psi_{1,2,3,4})$\! & \!$\nu(\scalq_{1,2,3},A_3)$\! \\ \hline
$\ell=1,\dots,k-1$ & $\ell+\frac{1}{2}$ & $\ell+1$ & $\ell+\frac{3}{2}$ \\
$\ell +1$ & $\ell+\frac{1}{2}$ & $-\ell$ & $\ell-\frac{1}{2}$ \\
$(k-1)(N-k)$ & $\frac{k}{2}$ & $\frac{k+1}{2}$ & $\frac{k+2}{2}$\\
$(k+1)(N-k)$ & $\frac{k}{2}$ & $-\frac{k-1}{2}$ & $\frac{k-2}{2}$ \\
$(N-k)(N-k)$ & $\frac{1}{2}$ & $0$ & $\frac{1}{2}$
\end{tabular}
\caption{\label{tab:spectrum}Masses and multiplicities of the different quantum fields, partially given in terms of $\nu$ defined in \eqref{eq: def nu} \cite{Buhl-Mortensen:2016pxs}. 
The ghost field $c$ arises from the gauge fixing.
Note that we have suppressed the $x_3$-dependence for ease of presentation.%
\protect\footnotemark
} 
\end{table}
\footnotetext{Note that the negative signs of the fermion masses can be absorbed into a chiral rotation of the fermions, cf.~\cite{Buhl-Mortensen:2016pxs}.}

The second non-standard property of the mass-like terms is that they depend on $1/x_3$, the inverse distance to the defect. 
This $x_3$-dependence in the mass terms can be exchanged for standard mass terms in $AdS_4$ \cite{Nagasaki:2011ue,Buhl-Mortensen:2016pxs}, which amounts to performing a Weyl transformation.
The scalar propagator resulting from this Weyl transformation can be extracted from the $AdS_4$ propagator given e.g.\ in \cite{Liu:1998ty}:
\begin{equation}
\label{eq: propagator}
\begin{aligned}
K(x,y)
=\frac{g_\YM^2\sqrt{x_3 y_3}}{2} \int \frac{\de^3 \vec{k}}{(2\pi)^3}\e^{i \vec{k}\cdot(\vec{x}-\vec{y})} I_{\nu}(|\vec{k}| x_3^<) K_{\nu}(|\vec{k}| x_3^>)\eqncom
\end{aligned}
\end{equation}
where $I_\nu$ and $K_\nu$ are modified Bessel functions and $x_3^<$ ($x_3^>$) is the smaller (larger) of $x_3$ and $y_3$. Moreover, $\vec{k}=(k_0,k_1,k_2)$ pertains to the directions parallel to the defect. 
The fermion propagator  can be extracted in a similar way but is not explicitly needed here.

\section{Wilson loop}
\label{sec: Wilson loop}

We now calculate the planar one-loop correction to a particular Wilson loop.
The colour and flavour parts of the required calculations are very similar to those appearing in the calculation of one-point functions, and we refer the reader to \cite{Buhl-Mortensen:2016pxs,longpaper} for details. 
The main difference lays in the space-time part.

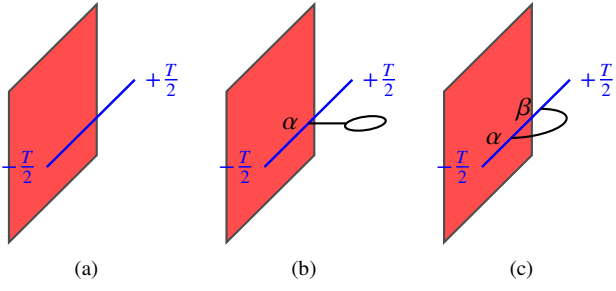
\begin{figure}[t]
\centering
 \subfigure[]{
\centering
\begin{tikzpicture}
	[	axis/.style={->,blue,thick},
		axisline/.style={blue,thick},
		cube/.style={opacity=.7, thick,fill=red}]
	\draw[cube] (0,-1,-1.5) -- (0,1,-1.5) -- (0,1,1.5) -- (0,-1,1.5) -- cycle;
\draw[axisline] (0.5,0,-1.5) -- (0.5,0,1.5) node[anchor=east]{$-\tfrac{T}{2}$};	
\draw[axisline] (0.5,0,1.5) -- (0.5,0,-1.5) node[anchor=west]{$+\tfrac{T}{2}$};
\end{tikzpicture}
  \label{subfig: tree}
} 
 \subfigure[]{
\centering
\begin{tikzpicture}
	[	axis/.style={->,blue,thick},
		axisline/.style={blue,thick},
		line/.style={black,thick},
		cube/.style={opacity=.7, thick,fill=red}]
	\draw[cube] (0,-1,-1.5) -- (0,1,-1.5) -- (0,1,1.5) -- (0,-1,1.5) -- cycle;
\draw[axisline] (0.5,0,-1.5) -- (0.5,0,1.5) node[anchor=east]{$-\tfrac{T}{2}$};	
\draw[axisline] (0.5,0,1.5) -- (0.5,0,-1.5) node[anchor=west]{$+\tfrac{T}{2}$};
\draw[line] (1,0,0) -- (0.5,0,0) node[anchor=east]{$\alpha$};
   \begin{scope}[canvas is xz plane at y=0]
     \draw[line] (1.25,0) circle (0.25);
   \end{scope}
\end{tikzpicture}
  \label{subfig: lollipop}
} 
 \subfigure[]{
\centering
\begin{tikzpicture}
	[	axis/.style={->,blue,thick},
		axisline/.style={blue,thick},
		line/.style={black,thick},
		cube/.style={opacity=.7, thick,fill=red}]
	\draw[cube] (0,-1,-1.5) -- (0,1,-1.5) -- (0,1,1.5) -- (0,-1,1.5) -- cycle;
\draw[axisline] (0.5,0,-1.5) -- (0.5,0,1.5) node[anchor=east]{$-\tfrac{T}{2}$};	
\draw[axisline] (0.5,0,1.5) -- (0.5,0,-1.5) node[anchor=west]{$+\tfrac{T}{2}$};
 \node[anchor=east] at (0.5,0,0.5) {$\alpha$};
 \node[anchor=east] at (0.5,0,-0.5) {$\beta$};
   \begin{scope}[canvas is xz plane at y=0]
     \draw[line] (0.5,0.5) arc (270:90:-0.5);
   \end{scope}
\end{tikzpicture}
  \label{subfig: tadpole}
}
\caption{Tree-level \subref{subfig: tree} and one-loop (\subref{subfig: lollipop} lollipop and \subref{subfig: tadpole} tadpole) contributions to the expectation value of the Wilson line. The defect is drawn in red, the classical Wilson line in blue and propagators in black.}
\label{fig: Wilson loop}
\end{figure}

\paragraph{Set-up}
Following \cite{Nagasaki:2011ue}, we consider a straight Wilson line parallel to the defect, which we can parametrise using 
$\gamma(\alpha)=\alpha n +(0,0,0,x_3)$ with $n$ being a unit vector with $n_3=0$. Without loss of generality, we take $n=(1,0,0,0)$.
Let 
\newcommand{\genA}{\mathcal{A}}
\newcommand{\genAc}{\genA^{\text{cl}}}
\newcommand{\genAq}{\tilde{\genA}}
\newcommand{\U}{U}
\newcommand{\Ucl}{\U^{\text{cl}}}
\begin{equation}
 \U(\alpha,\beta)=P \exp\int_{\alpha}^{\beta}\de t \genA(t)\,\,\,\text{with}\,\,\,\genA=iA_0-\sin\chi \scal_3-\cos\chi \scal_6
\end{equation}
be the parallel propagator along this line, where we have abbreviated $U(\alpha,\beta)\equiv U(\gamma(\alpha),\gamma(\beta))$, etc.
We then close the open colour indices and define 
\begin{equation}
 W=\tr \U(-\tfrac{T}{2},+\tfrac{T}{2})\eqndot
\end{equation}
In the limit $T\to\infty$, this yields an infinite straight Wilson line.
We depict the classical Wilson line along with its one-loop corrections in figure \ref{fig: Wilson loop}.

Expanding the field $\genA$ as
\begin{equation}
 \genA=\genAc+\genAq\eqncom
\end{equation}
the parallel propagator becomes 
\begin{align}
\label{eq: parallel propagator expanded}
  &\U(-\tfrac{T}{2},+\tfrac{T}{2})=\Ucl(-\tfrac{T}{2},+\tfrac{T}{2})\\\nonumber
   &+\int_{-\frac{T}{2}}^{+\frac{T}{2}}\de \alpha \Ucl(-\tfrac{T}{2},\alpha)\genAq(\alpha)\Ucl(\alpha,+\tfrac{T}{2})\\
   &+\int_{-\frac{T}{2}}^{+\frac{T}{2}}\de \alpha\int_{\alpha}^{+\frac{T}{2}}\de \beta \Ucl(-\tfrac{T}{2},\alpha)\genAq(\alpha)\Ucl(\alpha,\beta)\genAq(\beta)\Ucl(\beta,+\tfrac{T}{2})\nonumber\\
   &+\text{higher orders in the quantum fields}\eqncom \nonumber
 \end{align}
where $\Ucl$ denotes $\U$ with $\genA$ replaced by the classical field $\genAc$ and the higher orders in the quantum fields start to contribute only at two-loop order. 

\paragraph{Tree level}
At tree level, we have \cite{Nagasaki:2011ue}
\begin{equation}
\label{eq: tree-level Wilson loop}
 \begin{aligned}
  \langle W\rangle_{\text{tree}}&=\tr \Ucl(-\tfrac{T}{2},+\tfrac{T}{2})
  =\tr \exp\left(-T\sin\chi \scalc_3\right)\\
  &=(N-k)+\sum_{i=1}^k \exp\left(T\frac{\sin\chi}{x_3} d_{k,i}\right) \\
 & =(N-k)+\e^{-\frac{1}{2} (k-1) \frac{\sin\chi}{x_3}T} \frac{ 1-\e^{k \frac{\sin\chi}{x_3}T}}{1-\e^{\frac{\sin\chi}{x_3}T}}\\
  &\stackrel{{\mathclap{T\to\infty}}}{\simeq}\,\,\,\,\e^{\frac{1}{2} (k-1) \frac{\sin\chi}{x_3}T}
  \eqncom
 \end{aligned}
\end{equation}
where we have used that the only non-vanishing classical field in $\genA$ is $\scal_3$, which is diagonal with eigenvalues given in \eqref{eq: eigenvalues of t3}.

\paragraph{One loop}
At one-loop order, only two different diagrams contribute -- and both of them have direct counterparts in the calculation of one-point functions performed in \cite{Buhl-Mortensen:2016pxs}. The first diagram was called the lollipop diagram in \cite{Buhl-Mortensen:2016pxs} and is depicted in figure \ref{subfig: lollipop}, while the second diagram was called the tadpole diagram in \cite{Buhl-Mortensen:2016pxs} and is depicted in figure \ref{subfig: tadpole}. Although the second diagram no longer looks like a tadpole, we nevertheless keep the name.

\paragraph{Lollipop diagram}
The lollipop contribution stems from the second line in \eqref{eq: parallel propagator expanded}, where the quantum field is connected by a propagator to a cubic vertex in the action, whose other two fields are also connected by a propagator.
We find 
\begin{equation}
\begin{aligned}
\langle W\rangle_{\text{1-loop},\text{lol}}&=
\left\langle\tr\int_{-\frac{T}{2}}^{-\frac{T}{2}}\de \alpha\Ucl(-\tfrac{T}{2},\alpha)\genAq(\alpha)\Ucl(\alpha,+\tfrac{T}{2})\right\rangle\\
&= T \sum_i  \exp \left[ T \frac{\sin \chi}{x_3} d_{k,i} \right] \left\langle [\genAq]_{ii}\right\rangle_{\text{1-loop}} \eqncom
\end{aligned}
\end{equation}
where we have used that the parallel propagator is a diagonal matrix 
with non-zero entries
\begin{equation}
\label{eq: classical parallel propagator}
[\Ucl(\alpha,\beta)]_{ii} = \exp \left[ (\beta-\alpha) \frac{\sin \chi}{x_3} d_{k,i} \right]\eqndot
\end{equation}
In \cite{Buhl-Mortensen:2016pxs,longpaper}, we have found that, even at finite $N$, 
\begin{equation}
\begin{aligned}
 \langle [\genAq]_{ii}\rangle_{\text{1-loop}}&=0\eqncom
\end{aligned}
\end{equation}
when using a supersymmetry-preserving renormalisation scheme. 
Thus, the total lollipop contribution to the Wilson line also vanishes before taking the planar limit:
\begin{align}
 \langle W\rangle_{\text{1-loop},\text{lol}}&=0\eqndot
\label{eq: lolipop contribution to wilson loop}
\end{align}

\paragraph{Tadpole diagram}
In the tadpole diagram, the two quantum fields in the third line of \eqref{eq: parallel propagator expanded} are connected by a propagator.
In the large-$N$ limit, only the identity in either the parallel propagator in between the two quantum fields or in both other parallel propagators contributes.
Using \eqref{eq: classical parallel propagator}, we thus find that the planar contribution of the tadpole diagram is 
\begin{align}
\label{eq: tabpole contribution 1}
 &\langle W\rangle_{\text{1-loop},\text{tad}}=\\ \nonumber
 &\int_{-\frac{T}{2}}^{+\frac{T}{2}}\!\de \alpha\int_{\alpha}^{+\frac{T}{2}}\!\de \beta 
  \,\biggl(\exp \left[-(\alpha-\beta) \frac{\sin \chi}{x_3} d_{k,i}\right]
  \langle [\genAq]_{ai}(\alpha)[\genAq]_{ia}(\beta)\rangle\\\nonumber
 &\qquad\qquad +\exp \left[(\alpha-\beta+T) \frac{\sin \chi}{x_3} d_{k,i}\right]
  \langle [\genAq]_{ia}(\alpha)[\genAq]_{ai}(\beta)\rangle
  \biggr)\eqncom
 \end{align}
where $i=1,\ldots,k$ and $a = k+1,\ldots,N$ are summed over; note that the contribution of the $(N-k)\times(N-k)$ block vanishes.  
From \cite{Buhl-Mortensen:2016pxs,longpaper}, we know that in the large-$N$ limit
\begin{align}
  &\langle [\genAq]_{ia}(\alpha)[\genAq]_{aj}(\beta)\rangle
  =\langle [\genAq]_{ai}(\alpha)[\genAq]_{ja}(\beta)\rangle \\\nonumber 
  &=-\langle [A_0]_{ia}(\alpha)[A_0]_{aj}(\beta)\rangle
    +\sin^2\chi \langle [\scalq_3]_{ia}(\alpha)[\scalq_3]_{aj}(\beta)\rangle\\\nonumber
  &\hspace{0.25\textwidth}  
    +\cos^2\chi \langle [\scalq_6]_{ia}(\alpha)[\scalq_6]_{aj}(\beta)\rangle\\\nonumber
  &=\delta_{ij}N\sin^2\chi\left(
  \tfrac{k-1}{2k} K^{m^2=\frac{(k+2)^2-1}{4}} + \tfrac{k+1}{2k} K^{m^2=\frac{(k-2)^2-1}{4}} -K^{m^2=\frac{k^2-1}{4}} \right)\eqncom
 \end{align}
where the occurring propagators \eqref{eq: propagator} only depend on $\delta=\beta-\alpha$ and the distance $x_3$ to the defect. In particular, let us parameterise $x=(\alpha,0,0,x_3)$, $y=(\beta,0,0,0,x_3)$. Then \eqref{eq: propagator}  specialises to
\begin{equation}
K(x_3;\delta)
=\frac{g_\YM^2 x_3}{2} \int \frac{\de^3 \vec{k}}{(2\pi)^3}\e^{-i \vec{k}\cdot\vec{n}\, \delta} I_{\nu}(|\vec{k}| x_3) K_{\nu}(|\vec{k}| x_3)\eqndot
\end{equation}
In order to perform this integral, we decompose the $\vec{k}$ integration into spherical coordinates:
\begin{equation}
K(x_3;\delta)
=\frac{g_\YM^2 x_3}{8\pi^2} \int_0^\infty \!\de r\, r^2 \! \int_0^{\pi}\sin\theta \de\theta \e^{-i r \delta \cos\theta} I_{\nu}(r x_3) K_{\nu}(r x_3)\eqncom
\end{equation}
where we have already performed the trivial azimuth-angle integral.
The $\theta$ integration yields
\begin{equation}
K(x_3;\delta )
=\frac{g_\YM^2 x_3}{(2\pi)^2} \int_0^\infty \de r\, r \frac{\sin( \delta r)}{\delta} I_{\nu}(r x_3) K_{\nu}(r x_3)\eqndot
\end{equation}

Next, let us turn to the $\alpha,\beta$ integrations in \eqref{eq: tabpole contribution 1}. Since all functions in the integral only depend on $\delta$, we make the coordinate transformation $(\alpha,\beta)\rightarrow (\delta, \beta)$, for which the integral becomes
\begin{align}
\int_{-\frac{T}{2}}^{+\frac{T}{2}}\de \alpha\int_{\alpha}^{+\frac{T}{2}}\de \beta  \rightarrow 
\int_0^T\de \delta \int_{-\frac{T}{2}+\delta}^{+\frac{T}{2}}\de \beta \eqndot
\end{align}
Since the integrand only depends on $\delta$, we can trivially perform the $\beta$ integration resulting in a factor of $T-\delta$.

We are interested in the large-$T$ limit, which implies that from the prefactor in \eqref{eq: tabpole contribution 1} only the term with $i=1$ will contribute. For simplicity, let us introduce 
$\combi =   \frac{k-1}{2}\sin \chi$. Moreover, in order to make the $x_3$-dependence explicit, we rescale $r \rightarrow r/x_3$. This then combines into the following integral
\begin{align}
& \langle W\rangle_{\text{1-loop},\text{tad}} = \,
 \frac{\sin^2\chi}{x_3} \frac{\lambda}{(2\pi)^2}
 \int_0^T \de \delta \int_0^\infty \de r\, \nonumber\\
 &\, (T-\delta)\,  \biggl(\exp \left[\delta \, \combi/x_3 \right] + \exp \left[(T-\delta)\, \combi/x_3\right] \biggr)
 \frac{\sin ( \delta r/x_3)}{\delta} \\
 &\, r \left(
  \frac{k-1}{2k} I_{\frac{k+2}{2}}(r) K_{\frac{k+2}{2}}(r) +  \frac{k+1}{2k} I_{\frac{k-2}{2}}(r) K_{\frac{k-2}{2}}(r) - I_{\frac{k}{2}}(r) K_{\frac{k}{2}}(r) \right)\nonumber
  \eqndot
\end{align}
We can use partial integration on the Bessel function part including the factor of $r$. This eliminates the $\delta^{-1}$ term. As a result, the $\delta$ integral in the large-$T$ limit becomes straightforward. After making use of the Bessel function identities in \eqref{eq: Bessel function identities}, this finally leads us to the following integral:
\begin{align}
& \langle W\rangle_{\text{1-loop},\text{tad}} \stackrel{T\to\infty}{\simeq} \,
 \frac{\sin^2\chi}{x_3} \frac{\lambda}{4\pi^2} T \exp\left[\frac{\combi T}{x_3}\right]\\
 &\, \int_0^\infty \de r\,
\frac{\combi}{r^2+\combi^2} \left(\frac{1}{2} -
 r\, I^\prime_{\frac{k}{2} }(r) K_{\frac{k}{2}}(r)-
 \frac{1}{2} I_{\frac{k}{2} }(r) K_{\frac{k}{2}}(r)
\right)\eqndot\nonumber
\end{align}
The rational part can be easily integrated and we are finally left with
\begin{align}\label{eq:FinalIntegral}
& \langle W\rangle_{\text{1-loop},\text{tad}} \stackrel{T\to\infty}{\simeq} \,
 \frac{\sin^2\chi}{x_3} \frac{\lambda}{4\pi^2} T \exp\left[\frac{\combi T}{x_3}\right] \left(\frac{\pi}{4} - A\right) \eqncom\\
 \intertext{where}
 &\,  
A = \int_0^\infty \de r\,\frac{\combi}{r^2+\combi^2} \bigg[
 r\, I^\prime_{\frac{k}{2} }(r) K_{\frac{k}{2}}(r)+
 \frac{1}{2} I_{\frac{k}{2} }(r) K_{\frac{k}{2}}(r)
\bigg] \eqndot
\label{eq:AIntegral}
\end{align}

\paragraph{Odd $k$}

For odd values of $k$, the index on the Bessel functions becomes half-integer. For half-integer values, the Bessel functions $I_\nu,K_\nu$ are given by finite sums for which the integral \eqref{eq:AIntegral} can be carried out. The result is\footnote{
Note that the expression  is finite despite the negative argument of the hypergeometric function as $k$ is assumed to be odd.}
\begin{align}
&A= \frac{\pi}{4} - (\log (2 \combi )+\gamma_E )\frac{  _1F_2\left(\frac{3-k}{2};1-k,\frac{2-k}{2};-\combi ^2\right)}{2\combi^k}
\frac{\mathrm{\Gamma} \left(\frac{k}{2}\right) \mathrm{\Gamma} (k)}{\sqrt{\pi }\, \mathrm{\Gamma} \left(\frac{k-1}{2}\right)}\nonumber \\
&-\left(\frac{ i \combi
    K_{\frac{k-2}{2}}(i \combi ) + \frac{k-1}{2} K_{\frac{k}{2}}(i \combi )}{2 \pi\, i^k}K_{\frac{k}{2}}(i \combi ) (\text{Ei}(2 i \combi)-i\pi) + \text{c.c.}\right)
\label{eq: result A}    
    \\
&+
\sum_{l=0}^{\lfloor\frac{k+1}{4}\rfloor}\sum_{m=0}^{\frac{k-1}{2}}\sum_{n=1}^{2l+m}\frac{H_n}{(2l)! m! n!}
\frac{i^{k-1}(-1)^l}{(2\combi)^{2l+m-n}} \frac{\mathrm{\Gamma} \left(\frac{k+1}{2}+2 l\right) \mathrm{\Gamma} \left(\frac{k+1}{2}+m\right)}{\mathrm{\Gamma} \left(\frac{k+1}{2}-2 l\right) \mathrm{\Gamma} \left(\frac{k+1}{2}-m\right)}\nonumber\\
&\Bigg[\!\frac{[k^2 + (4 l-1)^2-2]\sin\! \frac{\pi (m+n)}{2} }{4 (4l+k-1)(4l-k-1)} - 
\frac{[k^2+ (4l+1)^2-2] \cos\! \frac{\pi  (m+n)}{2} }{32 (2 l+1) \combi }   \Bigg],\nonumber
\end{align}
where $\gamma_E$ is the Euler-Mascheroni constant, $H_n=\sum_{i=1}^n\frac{1}{i}$ is the harmonic number and $\mathrm{Ei}(y)$ is defined as the integral of $\e^{x}/x$ from $y$ to $\infty$. Notice also that when $k+1$ is divisible by $4$ the last term in the sum over $l$ has a spurious pole. The pole in the first term of the last line is cancelled by a zero from a Gamma function.

\paragraph{Large-$k$ limit}

The integral \eqref{eq:AIntegral}  can also be exactly evaluated when $k\rightarrow\infty$. In order to take the large-$k$ limit, we rescale the integration variable $r$ in \eqref{eq:AIntegral} by a factor of $k/2$. Recalling the definition of $\combi$ and using the asymptotic behaviour of the Bessel functions given in \eqref{eq: Bessel function asymptotics}, the integral can easily be performed. It yields
\begin{align}
\label{eq: large k Wilson loop}
\langle W\rangle_{\text{1-loop}}&= \langle W\rangle_{\text{1-loop},\text{tad}} \\
&\stackrel{\mathclap{T,k\to\infty}}{\simeq} \,\,\,
- \frac{T \exp\left[\frac{\combi T}{x_3}\right]}{x_3} \frac{\lambda}{8\pi^2k} 
\frac{\sin^2\chi}{\cos^3\chi}\biggl(\frac{\pi}{2}-\chi-\frac{1}{2}\sin2\chi\biggr)\eqndot
\nonumber
\end{align}

\paragraph{Particle-interface potential}

The expectation value of the Wilson loop is related to the particle-interface potential as
\begin{equation}
 \langle W(x_3)\rangle \cong \exp(-T\,V(x_3))\eqncom 
\end{equation}
for $T\to\infty$.
At tree level, we therefore have from \eqref{eq: tree-level Wilson loop}
\begin{equation}
 V_{\text{tree}}(x_3)=-\frac{k-1}{2x_3}\sin\chi\eqncom
\end{equation}
which agrees with \cite{Nagasaki:2011ue}.
At one-loop level, however, we expect 
\begin{equation}
 \langle W(x_3)\rangle_{\text{1-loop}} \cong -T\,V_{\text{1-loop}}(x_3)\exp(-T\,V_{\text{tree}}(x_3))\eqncom 
\end{equation}
such that we can read off the one-loop correction to the potential $V_{\text{1-loop}}(x_3)$ from the Wilson loop $\langle W(x_3)\rangle_{\text{1-loop}}$.

From the vanishing lollipop diagram \eqref{eq: lolipop contribution to wilson loop} and the
tadpole diagram \eqref{eq:FinalIntegral}, we then find the following contribution to the potential:
\begin{equation}
\label{eq: tadpole contribution to V 1-loop}
 V_{\text{1-loop}}(x_3)=V_{\text{1-loop,tad}}(x_3)=V_{\text{tree}}(x_3) \frac{\lambda}{2\pi^2} \frac{\sin \chi}{k-1}\left(\frac{\pi}{4} - A\right)  \eqndot
\end{equation} 

We notice that there is a point of enhanced symmetry for \mbox{$\chi=0$~\cite{Nagasaki:2011ue}}. At this point, the particle-interface potential
vanishes and correspondingly the expectation value of the Wilson line is equal to $N$.
The small $\chi=0$ expansion of~(\ref{eq: tadpole contribution to V 1-loop}) reads
\begin{align}
V_{\text{1-loop}}(x_3)= - \frac{\lambda}{2\pi^2}\frac{\chi^2}{x_3}\left[\frac{\pi}{8} \frac{k+2}{k}  + \frac{\chi}{k+1} +O(\chi^2)\right]\eqndot
\end{align}
The angle $\chi$ has some resemblance with the angle of the cusped Wilson loop in pure ${\cal N}=4$ SYM theory, for which the small angle expansion 
could be used to define a so-called Bremsstrahlung function related to the energy emitted by a moving quark~\cite{Correa:2012at, Fiol:2012sg}.
It would be interesting to further pursue this line of thought.

\section{Comparison to string theory}
 \label{sec: comparison}
 
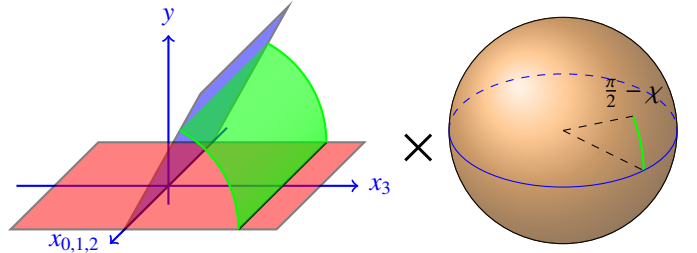
\begin{figure}[tb]
$
 \begin{aligned}
 \begin{tikzpicture}
 	[	D3/.style={opacity=.5,thick,fill=red},
 		D5/.style={opacity=.5,thick,fill=blue},
 		axis/.style={->,blue,thick},
 		string/.style={green,thick},
 		axisline/.style={blue,thick},
 		Wline/.style={black,thick},
 		cube/.style={opacity=.7, thick,fill=red}]

 	\draw[axisline] (-2,0,0) -- (0,0,0) node[anchor=west]{};	
			
 	\draw[axis] (0,0,0) -- (2.5,0,0) node[anchor=west]{$x_3$};
 	\draw[axis] (0,-0.25,0) -- (0,2,0) node[anchor=south]{$y$};
 	\draw[axis] (0,0,-2) -- (0,0,2) node[anchor=east]{$x_{0,1,2}$};

 	\draw[D5] (0,0,-1.5) -- (1,+1.8,-1.5) -- (1,1.8,1.5) -- (0,0,1.5) -- cycle;
 	
 	\draw[D3] (-1.5,0,-1.5) -- (2,0,-1.5) -- (2,0,1.5) -- (-1.5,0,1.5) -- cycle;
	
	\draw[Wline] (1.5,0,-1.5) -- (1.5,0,1.5) node[anchor=west]{};	
 	
\foreach \x in {-1.5,-1.49,...,1.5} 	{

     \tdplotdrawarc[string,opacity=0.15]{(0,0,\x)}{1.5}{61}{0}{}{};
}
     \tdplotdrawarc[string]{(0,0,-1.5)}{1.5}{61}{0}{}{};
     \tdplotdrawarc[string]{(0,0,1.5)}{1.5}{61}{0}{}{};
 
 \end{tikzpicture}
 \end{aligned} 
 \scalebox{2}{\raisebox{-0.3\baselineskip}{$\times$}}
 \,\,
\begin{aligned}
\tdplotsetmaincoords{60}{135}
\begin{tikzpicture}[scale=1.5,tdplot_main_coords]
\tikzstyle{string}=[thick,color=green,tdplot_rotated_coords]
     \tdplotsetrotatedcoords{0}{-90}{0}; 
    \draw[tdplot_main_coords] (0,0) circle (1cm);
    \draw[ball color=orange!80!white,opacity=0.69,tdplot_main_coords] (0,0) circle (1cm);
     \tdplotsetrotatedcoords{0}{-90}{0};
     \tdplotdrawarc[string]{(0,0,0)}{1}{60}{90}{}{};
      \draw[tdplot_main_coords,blue] (-1cm,0cm) arc (180:360:1cm and 0.5cm);
     \draw[dashed] (0,0,0) -- (0,1,0);
     \draw[dashed] (0,0,0) -- (0,0.86602540378,0.5) node[anchor=south] {$\frac{\pi}{2}-\chi$};
      \draw[dashed,tdplot_main_coords,blue] (-1cm,0cm) arc (180:0:1cm and 0.5cm);
\end{tikzpicture}
\end{aligned}
$
\label{fig: string theory}
\caption{The minimal surface corresponding to the Wilson loop. In the $AdS_5$ factor, the minimal surface (green) stretches from the Wilson loop (black) on the boundary (red) to the D5 brane (blue). In the $S^5$ factor, it is one-dimensional and stretches from the $S^2$ wrapped by the D5-brane (blue) to the latitude $\frac{\pi}{2}-\chi$ along constant longitude.}
\end{figure}

In the string-theory language, following the idea of~ \cite{Rey:1998ik,Maldacena:1998im, Drukker:1999zq},
the expectation value of the Wilson loop can be found in a
semi-classical limit, i.e.\ $N\rightarrow \infty$ followed by  $\lambda\rightarrow \infty$,  by evaluating the action of a classical 
string for which the worldsheet extends from the Wilson line in the boundary of $AdS_5$  to the D5 brane
in the interior 
and attaches to the D5 brane in such a way that the general D-brane boundary conditions
are fulfilled.  For this computation, which was done in~\cite{Nagasaki:2011ue}, the exact nature of the D5 brane
embedding is important. Let us write the metric of $AdS_5\times S^5$ as 
\begin{equation}
\de s^2= \frac{1}{y^2} \left(\de y^2+\de x^\mu \de x^\nu \eta_{\mu\nu}\right) +\de  \psi^2+\sin^2 \psi \de \Omega_2^2+\cos^2 \psi 
\de \tilde{\Omega}_2^2\eqncom
\end{equation}
where $x_\mu=(x_0,x_1,x_2)$, the boundary of $AdS_5$ is located at $y=0$ and
\begin{equation}
\de \Omega_2^2= \de \phi^2+\sin\phi^2 \de \theta^2\eqncom \hspace{0.5cm} \de \tilde{\Omega}_2^2= \de \tilde{\phi}^2+
\sin\tilde{\phi}^2 \de \tilde{\theta}^2\eqndot
\end{equation}
Furthermore, we assume that the background gauge field ${\cal F}$ has a flux on the untilded sphere, i.e.
\begin{equation}
{\cal F}\sim {k}\sin \theta\, \de \theta\, \de \phi \eqndot
\end{equation}
Then the world volume coordinates of the probe D5 brane are $(x_0,x_1,x_2, y, \theta, \phi)$ and its 
$AdS_4\times S^2$ embedding in 
$AdS_5\times S^5$ is described by $\tilde{\theta},\tilde{\phi}$ constant, $\psi=\frac{\pi}{2}$, and
\begin{equation}
y=\frac{\pi k}{\sqrt{\lambda}} \, x_3\eqndot
\end{equation}
In other words, the $AdS_4$ part of the D5 brane is tilted with respect to the $AdS_5$ boundary.  For the $AdS_4$ part 
of the D5 brane, the correct boundary conditions for the string equations of motion are of Neumann type along the brane and
of Dirichlet type transverse to the brane, meaning that the string must be perpendicular to the brane at the point of attachment.
This  boundary-value problem is of the same type as  for the classical pointlike string considered in~\cite{Buhl-Mortensen:2015gfd}, which is of relevance for the string-theory evaluation of one-point functions in the same defect set-up.
Let us parametrise the worldsheet using coordinates $(\tau=t,\sigma)$ with $t\in[-\infty,+\infty]$ and $\sigma\in [0,\sigma_1]$, where $\sigma=0$ corresponds to the end of the string which is attached to the $AdS_5$ boundary and
$\sigma_1$ corresponds to the end of the string which is attached to the D5 brane in the interior of $AdS_5\times S^5$.
The boundary conditions pertaining to the $S^5$ part of the background geometry then read~\cite{Nagasaki:2011ue}:
\begin{align}
\psi=\begin{cases}
      \chi\hspace{0.5cm}\mbox{for}\hspace{0.5cm} y=y(\sigma=0) =0\eqncom\\
      \frac{\pi}{2}\hspace{0.5cm}\mbox{for}\hspace{0.5cm} y=y(\sigma=\sigma_1)\eqndot
     \end{cases}
\end{align}
The solution of the classical string equations of motion with the above described boundary conditions can be uniquely
determined and the corresponding classical action evaluated~\cite{Nagasaki:2011ue}.  
We illustrate this in figure~\ref{fig: string theory}.
As usual in the string-theory evaluation of Wilson loops, the integral involved in the evaluation of the action has to be cut-off at a distance $\epsilon$ from the
boundary of AdS and the divergent $\frac{1}{\epsilon}$-piece removed before the result can be compared to a field-theory computation~\cite{Drukker:1999zq}. The authors of~\cite{Nagasaki:2011ue} found the particle-interface potential 
in closed form in the semi-classical limit $N\rightarrow \infty$ followed by $\lambda\rightarrow \infty$
but suggested to consider the further double-scaling limit
\begin{equation}
\lambda \rightarrow \infty, \hspace{0.5cm} k\rightarrow \infty, \hspace{0.5cm} \frac{\lambda}{k^2}\hspace{0.25cm} \mbox{ fixed},
\label{dsl}
\end{equation}
while keeping $k\ll N$. 
In this limit, the particle-interface potential reduces to~\cite{Nagasaki:2011ue}
\begin{equation}
 V=V_{\text{tree}}\left[1+\frac{\lambda}{4\pi^2k^2}\frac{\sin\chi}{\cos^3\chi}\left(\frac{\pi}{2}-\chi-\frac{1}{2}\sin2\chi\right) +\mathcal{O}\left(\frac{\lambda^2}{k^4}\right)\right]\eqndot \label{V}
\end{equation}
Taking  the double-scaling limit of our gauge-theory result obtained via \eqref{eq: large k Wilson loop}, we obtain perfect agreement with~(\ref{V}).

The double-scaling limit considered here is very reminiscent of the BMN limit, invented in connection with the
study of the spectral problem of ${\cal N}=4$ SYM theory, where 
another quantum number, $J$, with the
interpretation of an angular momentum was sent to infinity at the same time as $\lambda$ while the ratio
$\lambda/J^2$ was kept fixed~\cite{Berenstein:2002jq}.  In the case of the BMN limit, it eventually
turned out that starting at four-loop order the perturbative expansion of the gauge-theory anomalous dimensions did  not any longer  organise itself into a power series expansion in $\lambda/J^2$~\cite{Beisert:2006ez,Bern:2006ew,Cachazo:2006az}.
Nevertheless, the study of the BMN limit acted as a catalyst for the exploration of the integrability structure of the
AdS/CFT correspondence. Whether the gauge-theory observables of the defect set-up will continue to be well
defined in the limit~(\ref{dsl}) at higher loop orders  is an open question which
deserves further investigation. In any case, one could hope that the double-scaling limit~(\ref{dsl}) would be the catalyst for revealing the integrability
structure of the AdS/dCFT set-up.

\section{Conclusions \& Outlook}
\label{sec: conclusion}

In this letter, we have initiated the study of quantum corrections to non-local observables in a class of dCFTs with vevs, derived 
within the AdS/CFT set-up from  ${\cal N}=4$ SYM theory using the duality with certain probe-brane systems carrying background gauge field flux.  
Concretely, we have calculated the planar one-loop expectation value of an infinite straight Wilson line parallel to the defect, which allowed us
to infer the one-loop correction to the particle-interface potential in the dCFT.

Invoking the double-scaling limit described in the previous section, we have compared our result to the string-theory predictions of \cite{Nagasaki:2011ue} and found perfect agreement.
Considering the rather complicated structure of our result~\eqref{eq: large k Wilson loop}, 
the match we obtain is highly nontrivial. This result is furthermore in line with the results for one-point functions where the comparison of results between gauge and string theory likewise led to agreement~\cite{Buhl-Mortensen:2016pxs,longpaper}. 
Together, the two results thus provide a strong test of the gauge-gravity duality in the case where both conformal symmetry and supersymmetry are partially broken.

An extension of our results to finite $N$ and to two-loop order would be interesting and should in principle be doable, but it is rather technical and 
beyond the scope of the present publication.
It would likewise be interesting to generalise our one-loop analysis to more complicated Wilson loops. One example could be cusped Wilson lines, which have been extensively studied in $\mathcal{N}=4$ SYM theory where they yield among others the cusp anomalous dimension.  A cusped Wilson line would in the present set-up imply the introduction of further angles in addition to the angle of the cusp: the angles specifying the orientation of the cusp relative to the defect. Moreover, polygonal Wilson loops and their possible relation to scattering amplitudes in the dCFT might be worth exploring along the line of~\cite{Alday:2007hr,Brandhuber:2007yx,Drummond:2007cf}.  
In particular, the fate of the Yangian symmetry in the presence of a defect would be interesting to investigate for 
polygonal~\cite{Drummond:2008vq,Drummond:2009fd,Sever:2009aa,ArkaniHamed:2010kv,Beisert:2010gn} as well as for
smooth Wilson loops~\cite{Muller:2013rta}. 

Given that integrability has made its appearance in the study of one-point functions in the dCFT under consideration \cite{deLeeuw:2015hxa,Buhl-Mortensen:2015gfd,deLeeuw:2016umh}, it would be interesting to examine if integrability-based methods such as the quantum spectral curve could be applied in the calculation of certain Wilson loops as described for ${\cal N}=4$ in~\cite{Gromov:2015dfa,Gromov:2016rrp}. Finally, it would be interesting to investigate to which extent localisation methods,
which allow the exact evaluation of a sub-class of Wilson loops in ${\cal N}=4$ SYM theory (see e.g.\ \cite{Zarembo:2016bbk}), can be applied in the defect set-up.

\paragraph{Acknowledgements}

We are grateful to S.\ Caron-Huot, R.\ Janik, G.\ Semenoff and K.\ Zarembo for very helpful discussions.
M.d.L., C.K.\  and M.W.\ acknowledge partial support by  FNU  through
grants number DFF-1323-00082 and DFF-4002-00037. A.C.I.\ and, in parts, M.W.\ were supported by the ERC Advanced Grant 291092.

\appendix

\section{Bessel functions}\label{appBessel}

We use the following properties of Bessel functions, see for instance \cite{DLMF}:
\begin{equation}
\label{eq: Bessel function identities}
 \begin{aligned}
z \, I_\nu(z) &= 2\nu\, (I_{\nu+1}(z) - I_{\nu-1}(z)) \eqncom \\
2\,I^\prime_\nu(z) &= I_{\nu+1}(z) + I_{\nu-1}(z)\eqndot 
\end{aligned}
\end{equation}
Their asymptotic behaviour for $\nu\rightarrow\infty$ is
\begin{equation}
\label{eq: Bessel function asymptotics}
 \begin{aligned}
I_\nu(\nu z) &\sim \frac{\e^{\nu \xi}}{\zeta\sqrt{2\pi\nu}} \left[1 + \frac{1}{\nu}\left(\frac{3}{24\zeta}-\frac{5}{24\zeta^3}\right)+ \mathcal{O}(\nu^{-2})\right]\eqncom \\
K_\nu(\nu z) &\sim \frac{\pi \e^{-\nu \xi}}{\zeta\sqrt{2\pi\nu}} \left[1 - \frac{1}{\nu}\left(\frac{3}{24\zeta}-\frac{5}{24\zeta^3}\right)+ \mathcal{O}(\nu^{-2})\right]\eqncom \\
I^\prime_\nu(\nu z) &\sim \frac{\e^{\nu \xi}\zeta}{z\sqrt{2\pi\nu}} \left[1 - \frac{1}{\nu}\left(\frac{9}{24\zeta}-\frac{7}{24\zeta^3}\right)+ \mathcal{O}(\nu^{-2})\right]\eqncom
\end{aligned}
\end{equation}
where $\zeta = (1+z^2)^{1/4}$ and $\xi = \zeta^2 + \log\frac{z}{1+\zeta^2}$.

\section*{References}

\bibliographystyle{utphys2}
\bibliography{lettershortened}

\end{document}